# Magnetoresistance Anomalies in Ce-based Heavy Fermion Compounds

N E Sluchanko[1,2], A V Bogach[1,2], V V Glushkov[1,2], S V Demishev[1,2],

N A Samarin[1], G S Burhanov[3], O D Chistiakov[3] and D N Sluchanko[1]

[1] -A. M. Prokhorov General Physics Institute of RAS, 38, Vavilov str., Moscow, 119991, Russia.

[2] - Moscow Institute of Physics and Technology, 9, Institutskii per., Dolgoprudny, Moscow region, 141700, Russia.

[3] - A. A. Baikov Institute of Metallurgy and Materials Technology of RAS, 49, Leninskii pr., Moscow, 119991, Russia.

The work presents experimental results of precision magnetoresistance $\Delta\rho(H,T)$ measurements obtained for canonical heavy fermion compounds $CeAl_2$, $CeAl_3$, $CeCu_6$ and substitutional solid solutions $CeCu_{6-x}Au_x$ (x=0.1 and 0.2) and $Ce(Al_{0.95}M_{0.05})_2$ (M - Co, Ni). The research was performed in a wide range of temperatures (1.8-30K) and magnetic fields (up to 80 kOe). The data analysis indicates that the most consistent interpretation of magnetoresistance of both paramagnetic and magnetically ordered Ce-based systems with strong electron correlations can be obtained through the approach developed by K.Yosida (*Phys.Rev.,* **107**, *396(1957)*) that considers charge carrier scattering on localized magnetic moments in metallic matrix. Within this approach local magnetic susceptibility $\chi_{loc}(H,T_0) \equiv (1/H \cdot d(\Delta\rho/\rho)/dH)^{1/2}$ has been estimated directly from the magnetoresistance data $\Delta\rho/\rho = f(H,T_0)$. As a result, two additional contributions to magnetoresistance in Ce-based magnetic intermetallides have been established and classified. The procedure allowed to determine the peculiarities of magnetic phase H-T diagram as well as to reveal the electron density of states' renormalization effects in a wide vicinity of quantum critical point in the archetypal Ce-based systems with strong electron correlations.

**1.** Cerium based intermetallides are demonstrative examples of systems with strong electron correlations. These metallic heavy fermion compounds are characterized by a number of low-temperature anomalies in thermodynamic and transport characteristics. The features that may be especially singled out among them are (i) a significant Hall coefficient increase unusual to metallic systems [1, 2] and (ii) an abrupt decrease of resistivity in magnetic field [2].

The mechanism of large negative magnetoresistance (NM), observed in heavy fermion compounds at low temperatures, is commonly related to a suppression of the spin-flip scattering of conduction electrons on localized magnetic moments (LMM) of $Ce^{3+}$-ions with magnetic field. The NM effect is evidently established for the whole number of cerium intermetallides: so-called concentrated Kondo-systems with huge values of charge carrier mass ($CeCu_6$ [3], $CeAl_3$ [4]; $m^* \sim 200 \div 500 m_0$ [5], $m_0$- free electron mass), magnetic Kondo-lattices ($CeAl_2$ [2,6,7], $CeB_6$ [8], $CePb_3$ [9]), Kondo-insulators ($Ce_3Bi_4Pt_3$ [10], $CeNiSn$ [11]) and other compounds. An abrupt change of NM amplitude, which observed at metamagnetic transition in cerium compounds, is widely used in plotting of magnetic phase diagrams [2,12-13]. Thus the concordant, "correlated" behavior of magnetization and magnetoresistance in these strongly correlated electron systems (SCES) is commonly accepted. However, it is well known from the literature on this issue that there is a lack of the detailed quantitative analysis of NM effect in conductors in concentrated limit corresponding to the existence of Ce LMM in every unit cell of crystal lattice.

In this connection, the aim of the study was to perform precision measurements of magnetoresistance and to carry out the quantitative analysis of experimental results for cerium-based archetypal heavy fermion compounds and their substitutional solid solutions $CeAl_3$, $CeCu_{6-x}Au_x$ (x=0, 0.1, 0.2), $Ce(Al_{1-x}M_x)_2$ (x=0, 0.05, M=Co,Ni). Thus, the magnetoresistance was measured in the vicinity of quantum critical point (for $CeCu_{5.9}Au_{0.1}$ and $CeCu_{5.8}Au_{0.2}$ intermetallides [14-15]) as well as in heavy fermion regime (for canonical magnetic $CeAl_2$ and non-magnetic $CeCu_6$ and $CeAl_3$ systems). The undertaken investigation involved $CeAl_2$ and it's substitutional solid solutions $Ce(Al_{0.95}Co_{0.05})_2$ and $Ce(Al_{0.95}Ni_{0.05})_2$ to investigate the role of substitution disorder and doping effects of transition metal impurities (Co, Ni) on NM characteristics of cerium intermetallides.

**2.** The polycrystalline samples of the Ce-based intermetallic compounds were synthesized from stoichiometric quantities of high purity components in an electric arc furnace with a non-consumable tungsten crucible on copper water-cooled cold finger under a high purity helium atmosphere. The composition homogeneity was attained in the bulk materials by a repeated arc melting of the starting components in the stoichiometric ratio with subsequent annealing in evacuated quartz tubes. The samples' characterization methods (room temperature X-ray powder diffraction and electron microscopy) have been applied to control single-phase state of these compounds.

The measurements of transverse ($I \perp H$) magnetoresistance $\Delta\rho(T,H)$ of the above mentioned cerium intermetallides were performed using the dc-four-probe method in a wide range of temperatures 1.8-40K in magnetic fields up to 70 kOe. Two-channel nanovoltmeters Keithley 2182, included to the experimental setup similar to that one described in [16], were used for precision measurements of nanovolt level voltage from potential contacts to the sample. The required temperature stabilization accuracy (about 0.01K) of the measurement cell with the sample was achieved with the help of an original domestic temperature controller based on digital signal processors and standard resistance thermometer CERNOX 1050.

**3.** The resistivity temperature dependencies for intermetallides under investigation are shown in Fig.1. These $\rho(T)$ curves are in good consistence with the data previously obtained for $CeAl_2$ [2,6,7,16], $CeCu_6$ [3,17] and $CeAl_3$ [4,18] compounds. The features of the resistivity dependencies observed for aluminates $CeAl_2$ and $CeAl_3$ at temperatures near T~100K and T~40K, respectively, are usually associated with inelastic Kondo-scattering of conduction electrons on the LMM of excited levels of $Ce^{3+}$ $^2F_{5/2}$-state [2-4, 6-7, 16-18]. It is commonly believed, that the low-temperature maximum in $\rho(T)$ dependencies of Ce-based heavy fermion compounds (Fig.1a and inset b in Fig.1b) results from the crossover in scattering character of charge carriers from the spin-flip Kondo-scattering, which leads to LMM compensation on Ce-centers, to the coherent regime that realizes at low temperatures in dense Kondo-systems. As shown in Fig.1, the low temperature maximum of resistivity on $\rho(T)$ curves is observed in 2÷20K range for all heavy fermion compounds under investigation. In the special case of non-magnetic SCES $CeAl_3$ the low temperature maximum demonstrates a significantly lower amplitude in comparison with the aforementioned feature on $\rho(T)$ curve in the vicinity of T~40K.

The substitution of aluminium with nickel and cobalt in $Ce(Al_{1-x}M_x)_2$ system at x=0.05, and copper with gold in solid solutions $CeCu_{6-x}Au_x$ at x=0.1 and x=0.2 induces the noticeable resistivity increase (see Fig.1a and 1b). Simultaneously, the features on $\rho(T)$ dependency are blurring out with the advent of a substitutional disorder, and transition to coherent regime is suppressing. Note also that for $CeAl_2$ sample a kink connected with a transition to magnetically ordered (antiferromagnetic modulated, AFM ) state is observed in the vicinity of the low temperature maximum on $\rho(T)$ curve at $T_N \approx 3.85K$ (inset b in Fig.1b, curve 2).

Furthermore, for all investigated Ce-based heavy fermion compounds the measurements of magnetoresistance have been carried out at fixed temperatures in the range 1.8-30K. Typical dependencies of $\Delta\rho(H,T_0)/\rho_0 = (\rho(H,T_0) - \rho(0,T_0))/\rho(0,T_0)$ measured for two non-magnetic heavy fermion compounds $CeCu_6$ and $CeAl_3$ in magnetic fields up to 70 kOe are presented in Figs. 2a and 2b. Magnetoresistance curves for $CeCu_{6-x}Au_x$ (x=0.1, 0.2) intermetallides that also stay in paramagnetic state at T≥1.8K are basically similar to these ones presented in Fig.2. For these substitutional solid solutions the magnitude of NM effect in employed magnetic fields varies up to 13% (x=0.1) and 20% (x=0.2) respectively. Thus, the observed differences of $\Delta\rho(H,T_0)/\rho_0$ dependencies for $CeCu_{6-x}Au_x$ samples from the data in Fig.2 can be described only in terms of quantitative changes in NM amplitude.

Along with keeping the common trends of $\Delta\rho(H,T_0)/\rho_0$ curves presented in Fig.2, some significant differences in magnetoresistance behavior can be mentioned for magnetic intermetallides $CeAl_2$ (Fig.3), $Ce(Al_{0.95}Co_{0.05})_2$ (Fig.4a) and $Ce(Al_{0.95}Ni_{0.05})_2$ (Fig.4b) in the wide vicinity of magnetic phase transitions (see Fig.3b and insets in Fig.3-4). As shown in Figs.3-4, the NM effect reaching ~65% in $CeAl_2$ compound in magnetic fields up to 70 kOe and drops out to ~20% values in the case of aluminum substituted with cobalt and nickel in
$Ce(Al_{0.95}M_{0.05})_2$ (M=Co, Ni). The most significant feature on $\Delta\rho(H,T_0)/\rho_0$ curves for $CeAl_2$ is the abrupt resistivity change in the vicinity of transition from antiferromagnetic ($T<T_N\approx3.85K$) to paramagnetic phase ($T>T_N(H)$) (Fig.3b, see also [2,6-7,20-21]). In addition, positive magnetoresistance maxima for $CeAl_2$ (inset in Fig.3b T ≤ 3.5K) and $Ce(Al_{0.95}Co_{0.05})_2$ (inset in Fig.4a, T ≤ 4K) compounds are observed in field dependencies $\Delta\rho(H,T_0)/\rho_0$ in small enough magnetic fields $H<20$ kOe . The above features of low-temperature magnetoresistance seem to be connected with a rearrangement of complex magnetic-ordered state in $Ce(Al_{1-x}M_x)_2$ compounds with Laves-phase structure. Thus, these anomalies can be also used in the analysis of both magnetic H-T phase diagram and electronic states in these intermetallides.

The comparative quantitative analysis of contributions to the magnetoresistance for Ce-based heavy fermion compounds under investigation is presented in the next section.

**4.** Among various theoretical approaches applied to the interpretation of magnetoresistance anomalies in Ce-based intermetallides Kondo-impurity and Kondo-lattice models are preferred in most cases [22-27]. According to Zlatic calculations [22], a maximum of NM temperature dependence for $CeAl_3$ is connected with the Kondo temperature $T_K$ through the relation $T_{min}^{NM} \approx T_K/2$, whereas the transition of $\Delta\rho/\rho=f(T)$ from negative to positive should be observed at lower temperature $T_{inv} \approx T_K/2\pi$. In [23] the magnitude of NM effect for compounds with heavy fermions $CeAl_2$, $CeB_6$, $CeAl_3$ and $CeCu_2Si_2$ was calculated within the approach considering both Kondo-effect and crystal field splitting of $Ce^{3+}$ $^2F_{5/2}$-state. The nature of metamagnetic transition in SCES was analyzed within Kondo-lattice model in [24-25]. Additionally, the magnetoresistance effect connected with the metamagnetic transition was also estimated for $CeRu_2Si_2$ compound in [25]. Authors of [26-27] also used an approach which allows the suppression of $Ce^{3+}$ LMM Kondo-compensation mechanism when appears a local magnetization varying along the chains of $Ce^{3+}$ LMM at temperatures below $T_N$ in the matrix of $CeCu_{6-x}Au_x$ compounds with $x \geq 0.5$.

The aforementioned and other rare examples of analysis of the low-temperature magnetoresistance anomalies in cerium intermetallides are based on Kondo effect. The mechanism of the Kondo-compensation or the screening of Ce LMM through the spin-flip scattering of conduction electrons is accompanied by an emergence of many-body resonance in the electron density of states on the Fermi level $E_F$. Indeed, this interpretation can be considered as one of the available scenario of the appearance of heavy fermions in these compounds with strong electron correlations. However, this mechanism becomes inappropriate for explaining the origin and the features of nanosized magnetic regions, which were found in cerium intermetallides' matrix by various experimental methods at low temperatures (see, e.g., [28]).

An example of another approach to the interpretation of magnetoresistance anomalies in Ce-based systems can be found in [29] where the study of transport properties was performed both for paramagnetic and ferromagnetic phases of $CeRu_2Ge_2$. To explain the $\Delta\rho/\rho$ anomalies in $CeRu_2Ge_2$ the authors of [29] used the model of electron-magnon scattering. However, the approach applied in [29] explains neither the NM effect observed near Curie temperature nor the complex behavior with the inversion of $\Delta\rho/\rho$ sign in a wide temperature interval around the transition into ferromagnetic state.

In our point of view, one of the simplest, relatively long time known and widely applied approaches to the description of NM effect in conductors with embedded localized magnetic moments is the model proposed by K. Yosida in [19]. It was shown in [19] on the basis of calculations within s-d exchange model that conduction electron scattering on LMM leads to a rather large contribution to resistivity which is suppressed by external magnetic field. As a result, the NM effect proves to be proportional to squared local magnetization [19]

$$-\Delta\rho/\rho = 0.61 \langle M \rangle^2 / S^2 = \beta\, M_{loc}^2 \qquad (1)$$

and can be rewritten in a common case of several contributions to $M_{loc}$ as

$$-\Delta\rho/\rho = (\Sigma\beta M_{loc})^2 \qquad (1a).$$

For small enough magnetic fields, the relation (1) can be simplified as

$$-\Delta\rho/\rho = \beta\, \chi_{loc}^2\, H^2. \qquad (2)$$

Moreover, according to the predictions of [19], the maximum of NM effect has to correspond to the maximum of magnetic susceptibility $\chi(T)$ in the vicinity of Neel temperature $T_N$.

When applying the approach of [19] to the Ce-based heavy fermion systems it is necessary to take into account the nanosized magnetic domains (spin-polaron complexes) formed near the Ce-sites that has been recently established for CeAl$_2$ from the results of precision transport measurements [16]. In our opinion, the estimated values of localization radius of spin-polaron states ($a_p \approx 6\div 16$ Å [16]) are comparable to the typical spatial size $d_{sc}^m$ of charge carrier' magnetic scattering which is in order of the lattice constant of crystal structure - $d_{sc}^m \approx a \leq a_p \approx 10$ Å. Moreover, along with the main component $M_{loc}$ for Ce-based heavy fermion compounds with magnetic ordering an additional contribution to local magnetization $m_{loc}$ in combination with respective internal magnetic field $H_{int}$ should appear in the vicinity of 4f-centers in metallic matrix. For example, the value of $H_{int} \approx 75$ kOe was estimated for antiferromagnetic heavy fermion compound CeAl$_2$ from the analysis of low temperature polarized neutron diffraction in [30-32]. The similar value of $H_{int} \approx 79\pm 2$ kOe was also found in [33] from the analysis of CeAl$_2$ magnetostriction at helium temperatures. Thus, anticipating the results of magnetoresistance measurements and their analysis within (1)-(2) relations in Ce-based compounds with heavy fermions, the most complicated behavior $\Delta\rho/\rho=f(H,T)$ should be expected in the case of magnetically ordered cerium intermetallides.

Turning to the analysis of experimental results, it should be mentioned that the interpretation of the data of Fig.2-4 in terms of relations (1)-(2) allows to estimate the local magnetization $M_{loc}(T,H)$ in the immediate vicinity of Ce-centers in these SCES. According to (2), the numerical differentiation of magnetoresistance data (see, e.g., $d(\Delta\rho/\rho)/dH=f(H,T_0)$ curves in Fig.5a and 5b for $CeCu_{5.9}Au_{0.1}$ and $Ce(Al_{0.95}Co_{0.05})_2$ respectively) immediately results to the family of $\chi_{loc}(H,T_0) \equiv (1/H \cdot d(\Delta\rho/\rho)/dH)^{1/2}$ curves for all the intermetallides under investigation. Fig. 6-8 present the calculated values of $\chi_{loc}(H,T_0)$ for non-magnetic compounds with heavy fermions $CeCu_6$ and $CeAl_3$ (Fig.6), systems with the so-called "non-Fermi-liquid" or quantum critical behavior $CeCu_{5.9}Au_{0.1}$ and $CeCu_{5.8}Au_{0.2}$ (Fig.7) and magnetic solid solutions $Ce(Al_{0.95}Co_{0.05})_2$ and $Ce(Al_{0.95}Ni_{0.05})_2$ (Fig.8). As shown in Figs.6-8, the analysis of asymptotic behavior of $\chi_{loc}(H,T_0)$ provides rather rough estimation for saturation field $H_S$ of local magnetization in the systems under investigation. Thus, for all $CeCu_{6-x}Au_x$ compounds in the interval $0 \leq x \leq 0.2$ the value of $H_S$ is equal $180 \pm 20$ kOe, for $CeAl_3$ $H_S = 200 \pm 20$ kOe, and for substitutional solid solutions $Ce(Al_{0.95}M_{0.05})_2$ (M- Co,Ni) saturation field $H_S$ is $150 \pm 20$ kOe (see Figs. 6-8).

It should be point out that the phase diagram for $CeCu_{6-x}Au_x$ system with quantum critical point at $x \approx 0.1$ ($CeCu_{5.9}Au_{0.1}$, see upper panel in Fig.9) seems to be evidently established at present by numerous experimental methods (see [14-15, 34], and review article [35]). In this connection, a quantitative comparison of $\chi_{loc}(H,T_0 \leq 2K)$ results for different $CeCu_{6-x}Au_x$ compounds (Fig. 6-7) leads to the conclusion about the "reduction of magnetism" in the vicinity of quantum critical point (QCP in Fig.9). Figure 10 shows the temperature dependences $\chi_{loc}(T,H \to 0)$ for non-magnetic systems with heavy fermions $CeCu_{6-x}Au_x$ $0 \leq x \leq 0.2$ and $CeAl_3$, which were deduced directly from the low magnetic field data of Fig.6-7 (see also presentation $\chi_{loc}^{-1}(T)$ for $CeCu_{6-x}Au_x$ on the inset in Fig.10a). The $\chi_{loc}(T)$ behavior for $CeCu_6$ is in good agreement with the results of magnetic susceptibility measurement of [36] (see Fig. 10b). Moreover, a deviation of both local $\chi_{loc}(T)$ and bulk $\chi(T)$ susceptibilities from the Curie-Weiss type dependence

$$\chi_{loc}(T,H \to 0) = C/(T+\Theta_p) \qquad (3),$$

observed in T<8K temperature range (Fig.10b), seems to be associated with additional magnetic contribution into magnetic susceptibility of CeCu$_6$. A similar behavior of the $\chi_{loc}(T,H\rightarrow 0)$ parameter is also observed in CeAl$_3$ compound (see Fig.10c where the original data are compared to bulk magnetic susceptibility measurements taken from [37]). Besides, the value of paramagnetic Curie temperature $\Theta_p \approx -4.1\pm 0.1$K found for CeCu$_6$ in the interval T≥8K stays almost constant for all studied solid solutions with x from 0 to 0.2 in CeCu$_{6-x}$Au$_x$ family (see inset in Fig.10a).

Following to the approach of (1)-(2) the magnetoresistance data (Fig. 2-4) are directly connected to the local magnetization behaviour $M_{loc} \sim (-\Delta\rho/\rho)^{1/2}$. Therefore it is possible to estimate relative temperature changes of effective magnetic moment $\mu_{eff}(T)/\mu_{eff}(16K)=(C(T)/C(16K))^{1/2}$ when assuming $\Theta_p$ value to be constant and using the relation

$$M_{loc} \approx \chi_{loc}(0)H = C(T)H/(T+\Theta_p) \qquad (4)$$

for low magnetic fields interval. A rather good scaling of experimental curves in $(-\Delta\rho/\rho)^{1/2} = C(T)H/(T+\Theta_p)$ presentation for various temperatures $T_0$ (Fig.11) proves the correctness of the procedure giving the dependence of $\mu_{eff}(T)/\mu_{eff}(16K)$ (see inset in Fig.11). The significant increase (~20%) of the value $\mu_{eff}(T)$ in CeCu$_6$ in temperature range T≤10K does not have any explanation within the Kondo-lattice model. In fact, at lower temperatures the decrease of effective magnetic moment value on Ce-sites should be expected in the vicinity of Kondo temperature $T_K$ ($T_K$(CeCu$_6$)≈6K [38]) due to "switching on" of the LMM Kondo-compensation mechanism. On the other hand, the appearance of new magnetic contribution to $\chi_{loc}(0)$ for CeCu$_{6-x}$Au$_x$ compounds at T≤10K has a rather simple and natural explanation within spin-polaron approach [16] considering low temperature transformation of heavy fermions (spin polarons) into nanosized magnetic domains with an onset of coherent regime of spin fluctuations in the vicinity of Ce-centers.

As an alternative to the above scaling of the magnetoresistance data using Eq. (4) with $\Theta_p$=const (Fig.11), the appearance of the additional magnetic contribution to χ(T) can be described in terms of a variation of exchange field. In fact, the scaling of the magnetoresistance data with the help of the expression

$$(-\Delta\rho/\rho)^{1/2} = const \cdot H/(T+\Theta_p(T)) \qquad (5)$$

(see Fig.12) describes the local magnetization behavior with equally good accuracy for both $CeCu_6$ compound and its substitutional solid solutions $CeCu_{6-x}Au_x$ with x=0.1 and 0.2. The $\Theta_p(T)$ dependence resulting from Eq.(5) is presented in the inset in Fig.12. As shown in Fig.12, the quantum critical point x=0.1 is marked both by a drop of the normalized effective magnetic moment $\mu_{eff}(T)/\mu_{eff}(16K)$ as deduced from the low field slope of the $(-\Delta\rho/\rho)^{1/2} = f(H/(T+\Theta_p(T)))$ dependencies and by a significant decrease of $\Theta_p(T)$ parameter which determines the intensity of RKKI indirect exchange between LMM on Ce-sites (see dependencies of $\mu_{eff}(x)$ and $\Theta_p(x)$ in Fig.9). At the same time, at temperatures below 2K for $CeCu_{6-x}Au_x$ compounds with x=0.1 and 0.2 in the immediate vicinity of quantum critical point (QCP in Fig.9), one can observe significant deviations from the dependence (5) (Fig. 12). In our opinion, it may be explained in terms of abrupt changes in the character of quasi-particle interactions in the nearest neighborhood of QCP in these systems with strong electron correlations.

For $CeAl_3$ the scaling of Eq.(4) leads to $\Theta_p \approx -1.8 \pm 0.3K$ and, at temperatures below 7K, the normalized effective magnetic moment increase becomes comparable (~20%) to that one found for $CeCu_6$. Note that in the case of $CeAl_3$ the proximity of the first excited doublet of $Ce^{3+}$ $^2F_{5/2}$-state to the main one (crystal field splitting $\Delta_1 \approx 50K$ [39]) seems to be the reason of a significant quality degradation of magnetoresistance data scaling within relations (4) and (5).

When discussing the magnetoresistance data for magnetic compounds with Laves phase structure $Ce(Al_{1-x}M_x)_2$ (x=0 and 0.05, M=Co, Ni) note that these cerium intermetallides, having a magnetically ordered state at low temperatures, are characterized by appearing of additional features on field dependencies $\chi_{loc}(H,T_0)$ in the vicinity of $H_m \approx 11 \div 14$ kOe (see, e.g., curves for T≤9K on Fig.8a). In our opinion, to perform correctly a quantitative analysis of magnetoresistance data for these magnetic compounds within Eq. (1)-(2) it is necessary to separate a corresponding NM component, quadratic in low magnetic fields ($-\Delta\rho/\rho \sim H^2$, "Brillouin" contribution according to the terminology in [19]). The separation of contributions can be performed within the procedure where the derivative dependencies $d(\Delta\rho/\rho)/dH = f(H,T_0)$ are fitted by the linear dependence in the range of low magnetic fields $H \leq H_m \approx 11 \div 14$ kOe (see, e.g., Fig. 5b). In this approach the slope of linear fits of $d(\Delta\rho/\rho)/dH = f(H,T_0)$ determines $\chi_{loc}(0)$ parameter, whereas the cutoff on the ordinate axis (parameter A in Fig. 5b) corresponds to the additional linear $(-\Delta\rho/\rho \approx AH)$ contribution to the magnetoresistance of magnetic cerium intermetallides. Within this procedure, the experimental data of $\Delta\rho/\rho$ are divided into three components: (i) "Brillouin" contribution $-\Delta\rho/\rho \sim H^2$, (ii) linear term $-\Delta\rho/\rho \approx AH$ and (iii) low-field "magnetic" contribution $\Delta\rho/\rho|_{mag}$, obtained by the subtraction of the first two terms from the experimental curve (Fig. 13). Two families of curves $\chi_{loc}(H,T_0)$ for $CeAl_2$ and $Ce(Al_{0.95}Co_{0.05})_2$, derived from "Brillouin" magnetoresistance data within the offered procedure, are shown in Fig.14a and 14b respectively. Additionally, Fig.15 shows the low-field positive "magnetic" contributions to magnetoresistance $\Delta\rho/\rho|_{mag}$ for cerium intermetallides $Ce(Al_{0.95}Co_{0.05})_2$ and $Ce(Al_{0.95}Ni_{0.05})_2$ which seem to characterize the magnetization process and the changes of magnetic sub-structure with the increase of magnetic field in the interval $H \leq 14 kOe$. This positive "magnetic" contribution to magnetoresistance appears for $Ce(Al_{0.95}M_{0.05})_2$ compounds at temperatures $T<T^*\approx 15K$ and saturates at $H\approx H_m \approx 11\div 14$ kOe, whereas its amplitude reaches the maximum value at $T\approx 4\div 6K$ (see Fig.16a).

To describe all these components in magnetoresistance $\Delta\rho/\rho$ of magnetic Laves phases $Ce(Al_{0.95}M_{0.05})_2$, the relations (1)-(2) have to be expanded to include several additive components in magnetization $<M>$. A rather good quantitative description can be derived within relation:

$$-\Delta\rho/\rho = \beta(M_{loc}+m_{loc})^2 \qquad (6).$$

The additional term $m_{loc}$ in the right part of (6) corresponds to a small ferromagnetic contribution to the magnetization from nanosize magnetic domains appearing in the Laves phase matrix at temperatures $T<T^*$. Within this approach the positive contribution to $\Delta\rho/\rho$ (Fig.15) should result from the magnetization of the nanosize ferromagnetic domains. In this scenario the decrease of the $\Delta\rho/\rho|_{mag}$ amplitude at $T<3K$ (Fig.16a) seems to be due to an onset of the interaction between the magnetic nanoregions and a transition to long range magnetic ordering in $Ce(Al_{0.95}M_{0.05})_2$ intermetallic compounds.

At low magnetic fields expression (6) can be rewritten as:

$$-\Delta\rho/\rho \approx \beta \chi_{loc}^2 H^2 + 2\beta \chi_{loc} m_{loc} H + \beta m_{loc}^2 \qquad (7).$$

From the comparison between the coefficients in (7) and the amplitudes of negative "Brillouin", linear and positive "magnetic" contributions to magnetoresistance (see, e.g., Fig.13a and Figs.15-16) a conclusion can be drawn in favor of the proprietary use of relations (6)-(7). However, the summing rule of vectors $M_{loc}$ and $m_{loc}$, which determine the magnetoresistance behavior, should be further investigated in details.

Note that both negative linear and positive "magnetic" contributions appear due to the small ferromagnetic component $m_{loc}(T)$ in local magnetization (see. (6)-(7)) that is proved by the comparison of temperature dependences of these contributions to magnetoresistance (see Fig. 16). Besides, the metamagnetic transition (see, e.g., Fig.8a), which has been detected in magnetic field $H \approx H_m \approx 11 \div 14$ kOe at T<10K exclusively within the analysis of the $\Delta\rho/\rho|_{mag}$ contribution, should be treated as a feature of the $m_{loc}(T,H)$ field dependence, corresponding to the magnetization process of ferromagnetic nanodomains.

As it was suggested before, the behavior of the contributions to magnetoresistance for magnetic compound CeAl$_2$ in a wide vicinity of Neel temperature $T_N \approx 3.85$K (2K<$T$<5K) appears to be significantly more complicated then that of the other Ce-based intermetallides. At particular, a linear ($\Delta\rho/\rho \approx AH$) positive component $2\beta\chi_{loc}m_{loc}H$ and sign-reversal component $\Delta\rho/\rho|_{mag}$, which corresponds to the variation of $m_{loc}(T, H)$ and saturates in magnetic fields $H \approx H_m \approx 11 \div 14$ kOe (Fig. 13b), have been also detected in addition to the main Brillouin NM contribution. The temperature dependencies of the specified contributions to magnetoresistance of CeAl$_2$ (fig. 17) shows that the coefficient at linear contribution $2\beta\chi_{loc}m_{loc}$ is nonmonotonous by depends on temperature (Fig.17b), whereas the magnetic contribution $\Delta\rho/\rho|_{mag} = f(T,H)$ demonstrates the sign inversion in vicinity of $T \approx 3$K (see also Fig.18). Apparently, this complicated temperature variation of linear and magnetic components in the magnetoresistance of magnetic Laves phase CeAl$_2$ may be connected with a noticeable transformatoin of magnetic structure at temperature 2-5K.

It should be mentioned that the conclusion about the incommensurate sine-modulated structure ($K_I$ – type) of antiferromagnetic phase in *CeAl$_2$* made in [30] has been frequently discussed and criticized during recent decades. As an example, on the basis of neutron diffraction data measured on the single crystals of *CeAl$_2$* authors of [40] have concluded in favour of complicated antiferromagnetically modulated (AFM) state corresponding to the magnetic structure with three-component wave vector $k$ ($K_{III}$ – type) and 24-component order parameter (see also [41]). Later, authors of [42-43] presented arguments proving the origin of the *CeAl$_2$* magnetic structure as double elliptical spiral where the magnetic moments in two fcc Ce-sublattices are rotated in opposite directions. In this structure the magnitude of the magnetic moments varies slightly from site to site along every component of the spiral ($K_{II}$– type of magnetic structure). The variation in absolute value of $Ce^{3+}$ LMM along the spirals (so-called magnetic structure modulation), which is about 30% in the vicinity of $T_N$ [43], seems to be due to Kondo-compensation mechanism.

During last decade the magnetic structure of *CeAl₂* has been studied intensively with the help of μSR-spectroscopy technique [28, 44-45]. However, there is a substantial discrepancy in the results and conclusions of these investigations. One of the main reasons impeding the research of *CeAl₂* magnetic ground state, according to [28, 33, 46-47], might be a dependence of the transition temperature $T_N$ and of the long range magnetic ordering peculiarities on the inner local tensions and on the low concentrated impurities in the investigated samples. These factors lead both to the dispersion of Neel temperature value in the interval $T_N = 3.4 \div 3.9$ K [28], and, according to [46], to the realization of two magnetic phase transitions with close values of $T_N$, which result in the above mentioned problems in the interpretation of magnetic structure and features of *H-T* diagram in this compound. Recently, on the basis of the results of high precision charge transport measurements, it was shown that two magnetic phase transitions are really observed at temperatures $T_N = 3.85$ K and $T_{NI} = 3$K [21]. Moreover, the Hall coefficient anomalies found in [21] prove the conclusions of [14, 48-49] about the existence of ferromagnetic correlations in *CeAl₂* matrix at temperatures $T_N \leq T \leq 12K$.

Returning to the analysis of the magnetoresistance contributions in CeAl₂, it should be noted that the anomalies of magnetostriction and thermal expansion, observed in this compound at temperatures 2-5K in magnetic fields up to 15kOe [50-51] were interpreted by the authors in terms of changes of the antiferromagnetic domains polarization with a reorientation of the $Ce^{3+}$ LMM along the directions perpendicular to magnetic field **H**. However, taking into account the results of the present analysis of the magnetic contribution to magnetoresistance $\Delta\rho/\rho|_{mag}$, more reasonable explanation can be given from the consideration of the magnetization process in the ferromagnetic nanodomains. In this scenario the nanodomains, which are formed in CeAl₂ matrix in coherent regime of fast spin fluctuations near the Ce-sites, are responsible for the emergence of the small $m_{loc}(T, H)$ saturated in magnetic fields H~12-14kOe magnetization component and, as a result, for an enhancement of magnetic scattering $\Delta\rho/\rho|_{mag} \sim m_{loc}^2$ of charge carriers on these nanosize magnetic inhomogeneities. Thus, the appearance of nanosize ferromagnetic domains in the vicinity of Ce-sites can be considered as the reason for $\mu_{eff}$ to exceed the maximum value of magnetic moment of $Ce^{3+}$ ion ground state ($\Gamma_7$-doublet), observed in [49] at T ≤5K. It should be also pointed out that a complicated temperature behavior of contributions $\Delta\rho/\rho|_{mag}(H,T)$ and $\Delta\rho/\rho \approx A(T)H$ (see Fig. 17a and 17b) is directly connected to the magnetic phase transitions in $CeAl_2$ compound. As can be clearly seen from Fig.17, the inflection point of curves presented in Fig.17 corresponds to Neel temperature $T_N \approx 3.85K$, whereas features can be also observed at temperatures $T_N^* \approx 3.5K$ and $T_{NI} \approx 3K$. It is worth to note, that $T_N^* \approx 3.5K$ is usually associated with the anomalies in thermal capacity and thermal expansion [46], as well as with the features of the NQR characteristics [47] for $CeAl_2$ polycrystals at magnetic phase transition temperature. Besides, from the data of Fig. 17 one can conclude in favour of "correlated" behaviour of magnetic $\Delta\rho/\rho|_{mag}(H,T)$ and linear $\Delta\rho/\rho \approx A(T)H$ components in magnetoresistance in the temperature range of 3-5K.

However, for a detailed investigation of the nature of magnetic and linear contributions to magnetoresistance of this and others magnetic cerium intermetallides the detailed magnetic and transport characteristics measurements should be carried out on the single crystalline samples of these SCES.

It was shown in present study that the magnitude of magnetoresistance for all investigated SCES (Figs. 2-4) is basically determined by negative quadratic ($\sim H^2$) term in $\Delta\rho/\rho$ ("Brillouin" in terms of [19]). So, the analysis of this contribution to NM will be also presented below for $Ce(Al_{1-x}M_x)_2$ magnetic Laves phases. The data of Fig.19 present the temperature dependencies of $\chi_{loc}(T)$ and $\chi_{loc}^{-1}(T)$ parameters for all three studied magnetic compounds $Ce(Al_{1-x}M_x)_2$. As shown in Fig.19, a significant deviation of magnetic susceptibility from Curie-Weiss law (3) (dashed line in Fig.19) is observed for $CeAl_2$ in the vicinity of $T^* \approx 12K$. Note that the interval $T<T^*$ is characterized by appearance of a short range ferromagnetic correlations in the Ce-based Laves phase matrix [14, 48-49]. Similar behavior connected with a partial depression of the paramagnetic response has been also found below $T^* \approx 15K$ in $Ce(Al_{0.95}M_{0.05})_2$ (M-Co,Ni) compounds (Fig. 19). As a result, the Curie-Weiss behavior (3), which is characterized by the positive values of paramagnetic Curie temperature $\theta_p(CeAl_2) \approx 3.8K$ and $\theta_p(Ce(Al_{0.95}M_{0.05})_2) \approx 3.2 \div 3.5K$ at $T>T^*$, is changed to slower variation of $\chi_{loc}(T)$ parameter in the range $T<T^*$. The anomaly of $\chi_{loc}(T)$ in $CeAl_2$ observed at Neel temperature $T_N \approx 3.85K$ is comparable with the behavior of a bulk magnetic susceptibility $\chi(T)$ in this compound [48] (see Fig. 19). Similar to the case of $CeCu_6$ and $CeAl_3$ (Figs. 10b и 10c), $\chi_{loc}(T)$ and $\chi(T)$ in paramagnetic phase ($T \geq 5K$) are practically matched each other in $CeAl_2$, whereas significant differences between these susceptibility dependencies are observed in the vicinity of Neel temperature ($2K \leq T_N < 5K$) (see Fig.19a). In our point of view, these differences in $\chi_{loc}(T)$ and $\chi(T)$ should be attributed to the features of charge carrier scattering on the magnetic structure inhomogeneities, that evidently appear in the metallic matrix at temperatures near the magnetic phase transition.

Concerning the results presented in Fig.19, it should be noted that the depression of the paramagnetic response followed by the substitution of aluminum by cobalt or nickel leads to a significant decrease of $\chi_{loc}(T)$ value. Besides, the features near $T_m \approx 9K$ and 15K are clearly observed on the dependencies $\chi_{loc}(T)$ for substitutional solid solutions $Ce(Al_{0.95}M_{0.05})_2$ (Fig. 19). Moreover, as it was shown above, for temperature range T<10K the magnetic field dependencies $\chi_{loc}(T_0,H)$ demonstrate maximum, corresponding to the metamagnetic transition at $H_m \approx 11 \div 14$ kOe (see, e.g., curves on Fig.8a). So, the anomaly of $\chi_{loc}(T_0, H \to 0)$ at $T_m \approx 9K$ should be connected with the transition to the coherent regime of spin fluctuations and to the formation of the nanosize ferromagnetic domains in vicinity of Ce-sites in $CeM_2$ matrix.

Finally, the results of scaling, which is analogous to that one performed above for nonmagnetic SCES $CeCu_{6-x}Au_x$ and $CeAl_3$ (Figs.11-12), should be also given here for heavy fermions compounds $Ce(Al_{1-x}M_x)_2$. The analysis is based on the scaling of NM curves within the relation (5) with temperature dependent $\theta_p(T)$ parameter. The dependencies $(-\Delta\rho/\rho)^{1/2}=f(H/(T+\theta_p))$, obtained with the help of proposed scaling procedure, are shown in Fig.20, the deduced $\theta_p(T)$ temperature dependence being presented in Fig.21. It should be mentioned, that in paramagnetic phase the curve families $(-\Delta\rho/\rho)^{1/2}=f(H/(T+\theta_p))$ are very similar to each other with close in absolute value local magnetization $M_{loc}=(-\Delta\rho/\rho)^{1/2}$. At the same time, as Fig.20 shows, the analysis indicate significant degradation of the scaling quality in the interval of large values $H/(T+\theta_p)$, that, apparently, should be associated with essential suppression of antiferromagnetic interaction accompanied by the increase of local magnetization in solid solutions $(Al_{0.95}M_{0.05})_2$ in high magnetic field. Within this approach one can explain the abrupt increase of $M_{loc}(H/(T+\theta_p))$ with magnetic field in $CeAl_2$ (see Fig.20) which observed at the transition from antiferromagnetic modulated phase to the collinear ($M||H$) one in this extraordinary compound.

The analysis of $\theta_p(T)$ parameter in $Ce(Al_{1-x}M_x)_2$ compounds (Fig.21), obtained by scaling within relation (3), leads to the conclusion that there is a significant variation of quantitative characteristics of exchange interactions $\theta_p(T)$ (or $\mu_{eff}(T)$) with temperature in these systems. Thereby, the $\theta_p(T)$ temperature dependencies also have anomalies in the immediate vicinity of magnetic transitions at $T_N$ and $T_m$, caused by magnetic structure transformation and associated features of charge carrier scattering in magnetic intermetallides of cerium.

### 5. Conclusion.

The study presents the results of high precision measurements of magnetoresistance for several cerium intermetallides – nonmagnetic compounds with heavy fermions $CeCu_{6-x}Au_x$ (x=0, x=0.1, x=0.2), $CeAl_3$ and magnetic substitutional solid solutions $Ce(Al_{1-x}M_x)_2$ (x=0, x=0.05, M-Co, Ni). The experimental data have been obtained at low temperatures 1.8-30K in magnetic fields up to 70 kOe. The quantitative analysis allows to separate the contributions to magnetoresistance into the main "Brillouin" -$\Delta\rho/\rho \sim H^2$, linear $\Delta\rho/\rho \approx A(T)H$ and magnetic $\Delta\rho/\rho|_{mag}(H,T)$ components in $\Delta\rho/\rho$ as well. The appearance of the linear and magnetic contributions to magnetoresistance is associated with the peculiarities of magnetically ordered state at low temperatures and seems to be a characteristic feature of these Ce-based magnetic heavy fermion compounds. It was also shown that the most complete and adequate interpretation of magnetoresistance effect in studied intermetallides can be probably obtained in the frameworks of K. Yosida's model [19] that considering charge carrier scattering on localized magnetic moments embedded into metallic matrix. The comparison of the local magnetization and local susceptibility dependencies, obtained within the relation $\Delta\rho/\rho \sim M^2 \sim \chi^2 H^2$, with bulk magnetic characteristics of Ce-based heavy fermion compounds evidently proves the applicability of the offered approach to the description of the charge carrier transport in these SCES. The restrictions in application of the approaches based on Kondo-lattice model to the interpretation of the magnetoresistance of cerium intermetallides are discussed.

This work was supported by Russian Foundation for Basic Research (projects 04-02-16721, 05-08-33463), INTAS (no. 03-51-3036), the Program "Strongly Correlated Electrons in Semiconductors, Metals, and Magnetic Materials" of RAS and the Program "Development of Scientific Potential of High School" of MES RF (project 8145). V.V.G. acknowledges the financial support from Russian Science Support Foundation.

**REFERENCES.**

[1] P.Coleman, P.W.Anderson, T.V.Ramakrishnan, Phys. Rev. Lett., **55** (1985) 414

[2] F.Lapierre, P.Haen, A.Briggs et. al., J. Magn.Magn.Mat., **63-64**(1987) 76.

[3] T.Penney, F.P.Milliken, S. von Molnar et. al., Phys.Rev.B, **34** (1986) 5959.

[4] N.B.Brandt, V.V.Moschalkov, N.E.Sluchanko et al., Sov. Solid State Phys., **26** (1984) 913 in Russian.

[5] W.P.Beyerman, A.M.Awasthi, J.P.Carini, G.Gruner, J.Magn.Magn.Mat., **76-77**(1988)207.

[6] N.B.Brandt, V.V.Moschalkov, N.E.Sluchanko et al., Sov. Solid State Phys., **10** (1984) 940 in Russian.

[7] V.V.Moshchalkov, P.Coleridge, E.Fawcett, A.Sachrajda, Sol.St.Comm., **60** (1986) 893.

[8] A.Takase, K.Kojima, T.Komatsubara et. al., Sol.St.Comm., **36** (1980) 461.

[9] U.Welp, P.Haen, G.Bruls et al., J. Magn.Magn.Mat., **63-64** (1987) 28.

[10] G.S.Boebinger, A.Passner, P.C.Canfield, Z.Fisk, Physica B, **211** (1995) 227.

[11] T.Takabatake, M.Nigasawa, H.Fujii et. al., J. Magn.Magn.Mat., **108** (1992) 155.

[12] A.K.Nigam, S.Radha, S.B.Roy, G.Chandra, Physica B, **205** (1995) 421.

[13] H.P.Kunkel, X.Z.Zhou, P.A.Stampe, J.A.Cowen, G.Williams, Phys.Rev.B, **53** (1996) 15099.

[14] C.M.Varma, Z.Nussinov, W.van Saarloos, Phys.Rep., **361** (2002) 267.

[15] G.R.Stewart, Rev.Mod.Phys., **73** (2001) 797.

[16] N.E.Sluchanko, A.V.Bogach, V.V.Glushkov et al., JETP Lett., **76** (2002) 26.

[17] M.R.Lees, B.R.Coles, E.Bauer, N.Pillmayr, J.Phys.Cond.Mat., **2** (1990) 6403.

[18] Z.Fisk, H.R.Ott, T.M.Rice, J.L.Smith, Nature, **320** (1986) 124.

[19] K.Yosida, Phys. Rev., **107** (1957) 396.

[20] N.E.Sluchanko, A.V.Bogach, I.B.Voskoboynikov et al., Phys. Solid State, **45** (2003) 1096.

[21] N.E.Sluchanko, A.V.Bogach, V.V.Glushkov et al., JETP, **98** (2004) 793.

[22] V.Zlatic, J.Phys.F, **11** (1981) 2147.

[23] Y.Lassailly, A.K.Bhattacharjee, B.Coqblin, Phys. Rev. B, **31** (1985) 7424.

[24] R.Citro, A.Romano, J.Spalek, Physica B, **259-261** (1999) 213.

[25] T.M.Hong, G.A.Gehring, Phys.Rev.B, **46** (1992) 231.

[26]  H.von Lohneysen, A.Neubert, T.Pietrus et.al., Eur. Phys. J. B, **5** (1998) 447.

[27]  A.Rosch, P.Wolfle, A.Neubert et al., Physica B, **259-261** (1999) 385.

[28]  A.Amato, Rev.Mod.Phys., **69** (1997) 1119.

[29]  M.B.Fontes, S.L.Bud'ko, M.A.Continentino, E.M.Baggio-Saitovich, Physica B, **270** (1999) 255.

[30]  B.Barbara, J.X.Boucherle, J.L.Buevoz et al., Sol. St. Comm., **24**, 481 (1977).

[31]  B.Barbara, M.F.Rossignol, J.X.Boucherle et. al., Phys. Rev. Lett., **45**, 938 (1980)

[32]  A. Benoit, J.X.Boucherle, J.Flouquet et al., Valence Fluctuations in Solids, eds. by L.M.Falicov, W.Hanke, M,B,Maple, North-Holland Publ. Comp., 197-206 (1981).

[33]  E.Fawcett, V.Pluzhnikov, H.Klimker, Phys. Rev. B, **43**, 8531 (1991).

[34]  Q. Si, cond-mat/0302110v1.

[35]  C.M. Stishov, Uspekhi Fiz. Nauk, **174** (2004) 853 in Russian.

[36]  Y.Onuki, R.Settai, K.Sugiyama et al., J.Phys.Soc.Jpn.,**73** (2004) 769.

[37]  E.Borchi, S.De Gennaro, C.Taddei, Phys.Rev.B, **15** (1977) 4528.

[38]  H.von Lohneysen, M.Sieck, O.Stockert, M.Waffenschmidt, Physica B, **223-224** (1996) 471.

[39]  P.A.Alexeyev, I.P.Sadikov, I.A.Markova et al., Sov. Solid State Phys., **18** (1976) 2509 in Russian.

[40]  S.M. Schapiro, E.Gurewitz, R.D.Parks, L.C. Kupferberg, Phys. Rev. Lett., **43** (1979) 1748.

[41]  M.Ma, J. Solyom, Phys. Rev. B, **21** (1980) 5262.

[42]  E.M. Forgan, B.D. Rainford, S.L. Lee, J.S. Abell, Y.Bi, J. Phys. Cond. Mat., **2** (1990) 10211.

[43]  F.Giford, J.Schweizer, F.Tasset, Physica B, **234-236** (1997) 685.

[44]  A.Schenk, D.Andreica, M.Pinkpank, F.N.Gygax, H.R. Ott, A.Amato, R.H. Heffner, D.E. MacLaughlin, G.J. Nieuwenhuys, Physica B, **259-261** (1999) 14.

[45]  A.Schenk, D.Andreica, F.N.Gygax, H.R. Ott, Phys. Rev. B, **65** (2001) 024444.

[46]  R.Schefzyk, W.Lieke, F.Steglich, Sol.St.Comm., **54** (1985) 525.

[47]  J.L.Gavilano, J.Hunziker, O.Hodak, T.Sleator, F.Hulliger, H.R.Ott, Phys.Rev.B, **47** (1993) 3438.

[48]  M.C.Croft, R.P. Guertin, L.C. Kupferberg, R.D. Parks, Phys. Rev. B, **20** (1979) 2073.

[49]  F.Steglich, C.D.Bredl, M.Loewenhaupt, K.D.Schotte, J.Phys.Coll., **40** (1979) C5-301.

[50]     M.Croft, I.Zoric, R.D.Parks, Phys. Rev. B, **18** (1978) 345.

[51]     M.Croft, I.Zoric, R.D.Parks, Phys. Rev. B, **18** (1978) 5065.

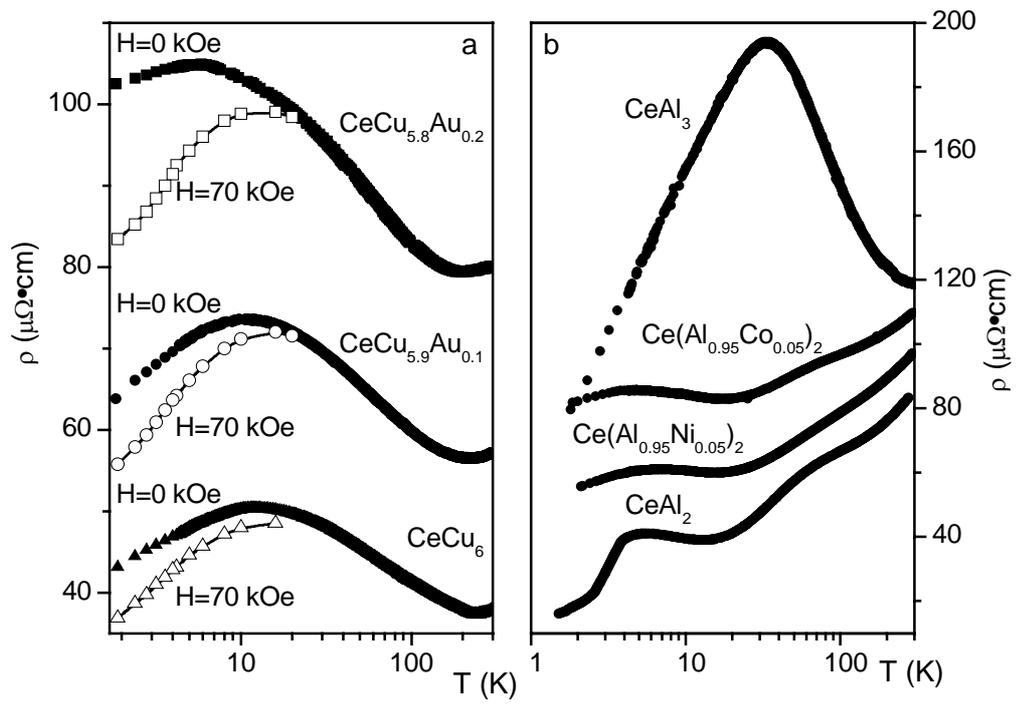

**Fig.1.** Temperature dependencies of resistivity $\rho(T)$ CeCu$_{6-x}$Au$_x$ (x=0, 0.1, 0.2) for various values of magnetic field ($H$ =0 kOe, 70 kOe) (**a**) and CeAl$_3$, CeAl$_2$, Ce(Al$_{0.95}$Ni$_{0.05}$)$_2$ and Ce(Al$_{0.95}$Co$_{0.05}$)$_2$ (**b**).

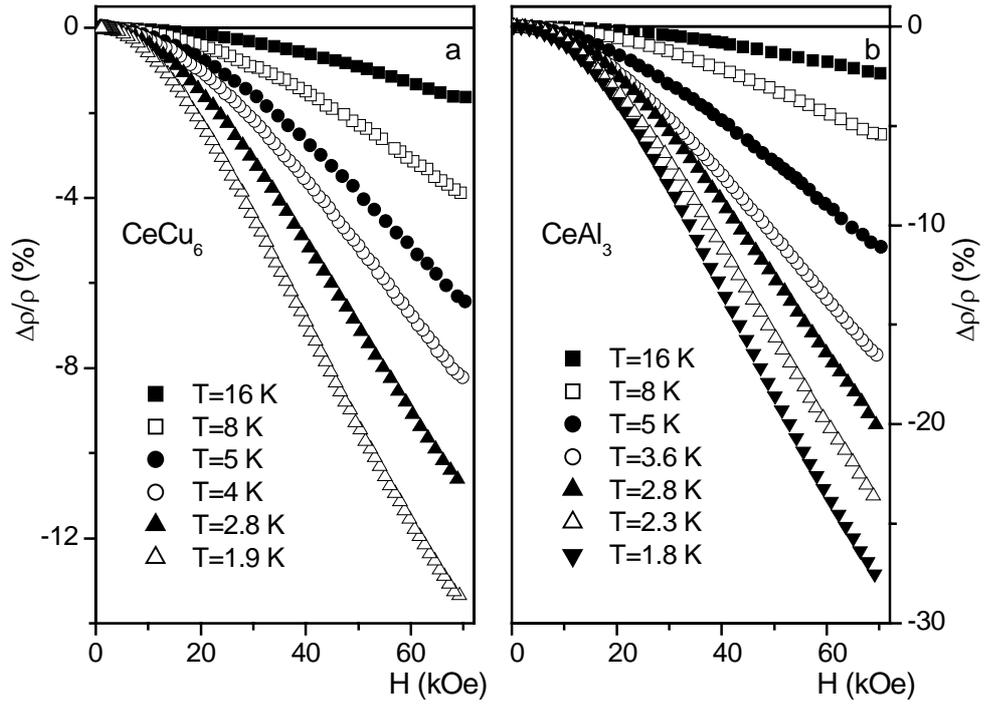

**Fig.2.** Field dependencies of magnetoresistance $\Delta\rho/\rho(H)$ CeCu$_6$ (**a**) and CeAl$_3$ (**b**) for temperatures $1.8\ \text{K} \leq T \leq 16\ \text{K}$.

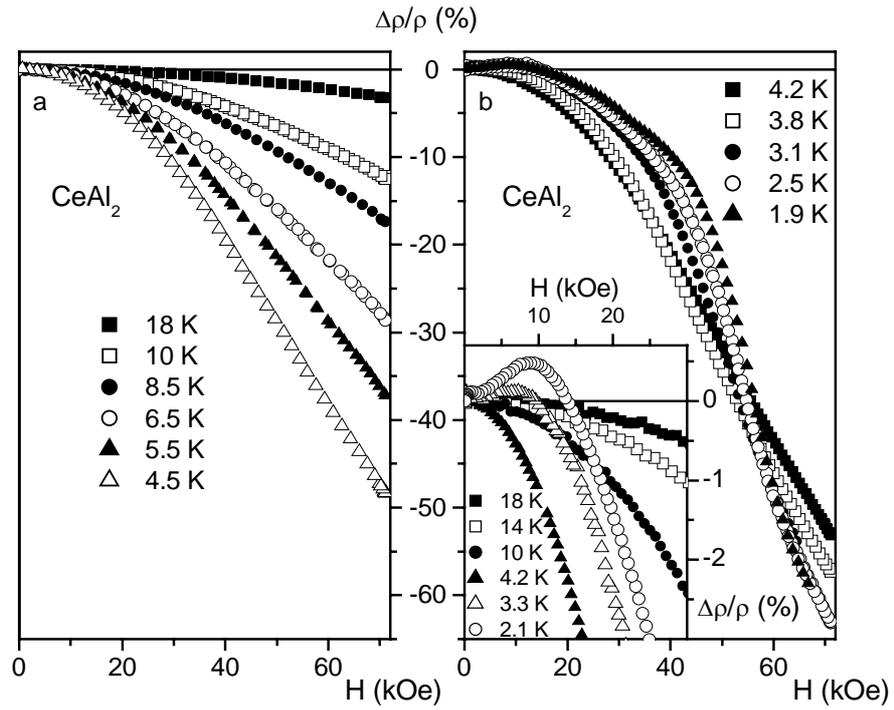

**Fig.3.** Field dependencies of magnetoresistance $\Delta\rho/\rho(H)$ CeAl$_2$ for temperatures $4.2\,\text{K} \leq T \leq 18\,\text{K}$ (**a**) and $1.9\,\text{K} \leq T \leq 4.2\,\text{K}$ (**b**). The inset presents low magnetic field magnetoresistance $\Delta\rho/\rho$ for CeAl$_2$ in expanded scale ($H < 30$ kOe).

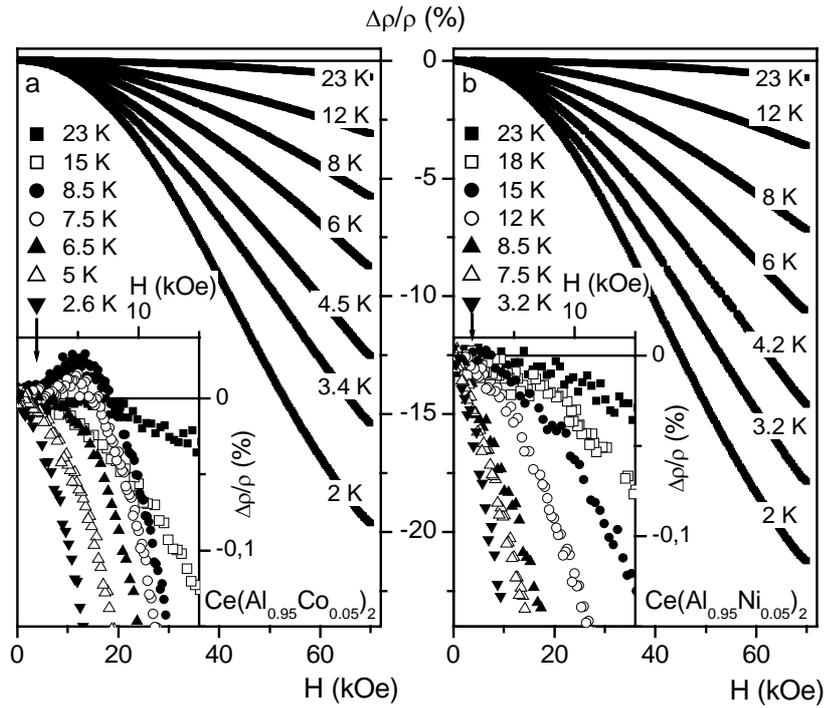

**Fig.4.** Field dependencies of magnetoresistance $\Delta\rho/\rho(H)$ Ce(Al$_{0.95}$Co$_{0.05}$)$_2$ (**a**) and Ce(Al$_{0.95}$Ni$_{0.05}$)$_2$ (**b**) for various temperatures. The insets present the magnetoresistance $\Delta\rho/\rho$ in small magnetic field area ($H$ < 15 kOe).

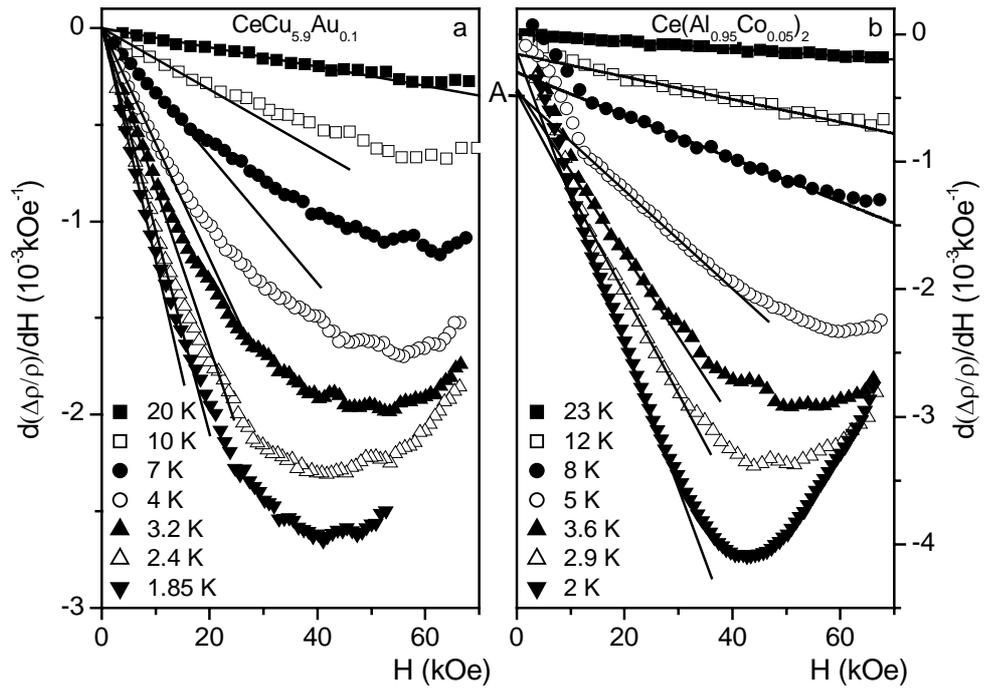

**Fig.5.** Field dependencies of magnetoresistance derivatives $d(\Delta\rho/\rho)/dH$ CeCu$_{5.9}$Au$_{0.1}$ (**a**) and Ce(Al$_{0.95}$Co$_{0.05}$)$_2$ (**b**) in various temperatures.

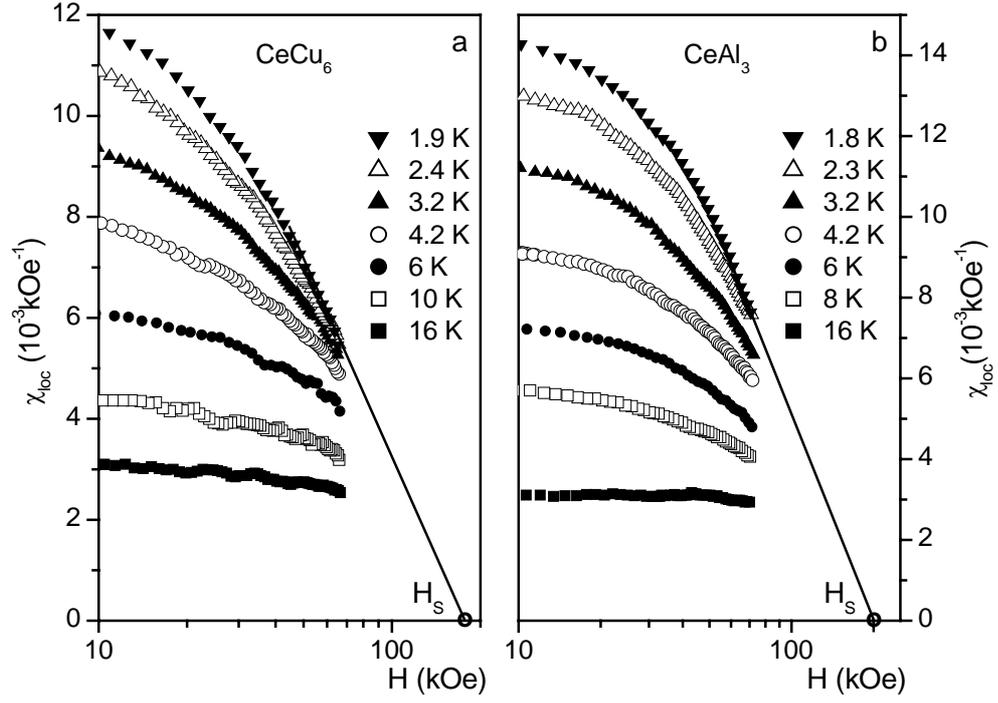

**Fig.6.** Field dependencies of local magnetic susceptibility $\chi_{loc}(H,T_0) \equiv (1/H \cdot d(\Delta\rho/\rho)/dH)^{1/2}$ (see text) for CeCu$_6$ (**a**) and CeAl$_3$ (**b**) in temperatures 1.8 K $\leq T \leq$ 16 K. Dots on the X-axis indicate magnetic field values $H_s$, corresponding to saturation of local magnetization (see text) $H_s \approx$ 180 kOe for CeCu$_6$ and $H_s \approx$ 200 kOe for CeAl$_3$.

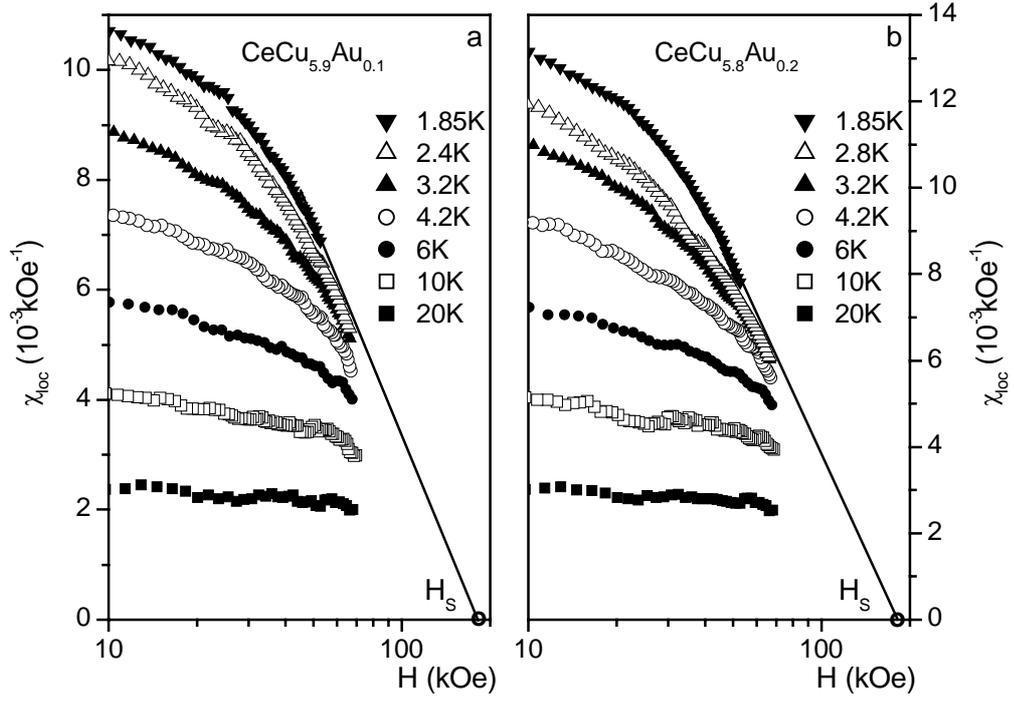

**Fig.7.** Field dependencies of local magnetic susceptibility $\chi_{loc}(H,T_0)$ CeCu$_{5.9}$Au$_{0.1}$ (**a**) and CeCu$_{5.8}$Au$_{0.2}$ (**b**) for temperatures 1.85 K $\leq T \leq$ 20 K. Dots on the X-axis indicate magnetic field values $H_s$, corresponding to saturation of local magnetization $H_s \approx$ 180 kOe.

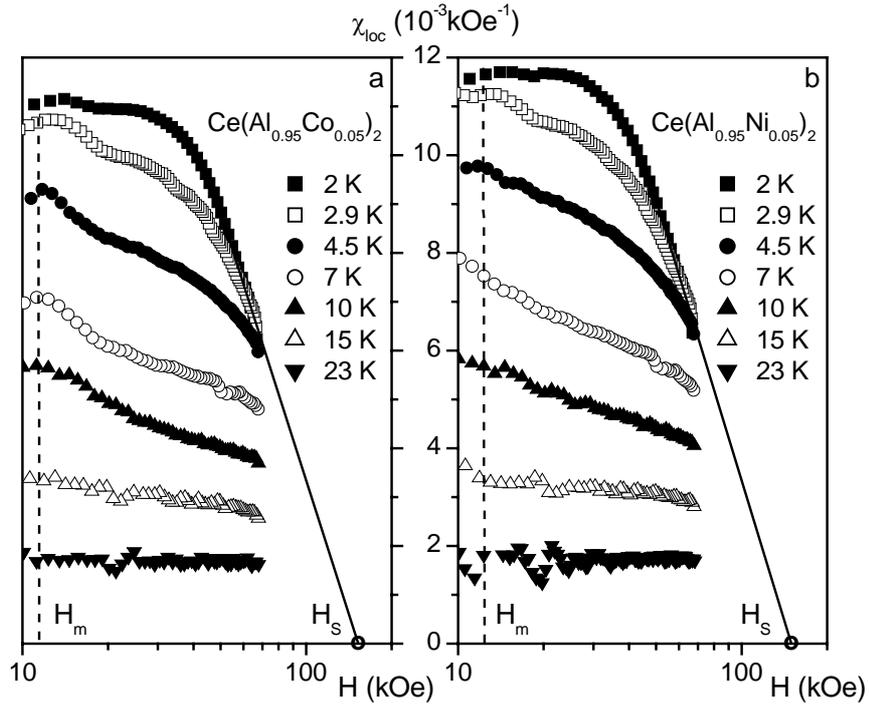

**Fig.8.** Field dependencies of local magnetic susceptibility $\chi_{loc}(H,T_0)$ Ce(Al$_{0.95}$Co$_{0.05}$)$_2$ (**a**) and Ce(Al$_{0.95}$Ni$_{0.05}$)$_2$ (**b**) for temperatures 2 K $\leq T \leq$ 23 K. Dots on X-axis indicate magnetic field values $H_s$, corresponding to saturation of local magnetization $H_s \approx$ 150 kOe. $H_m \approx$ 11 kOe for Ce(Al$_{0.95}$Co$_{0.05}$)$_2$ and $H_m \approx$ 12.5 kOe for Ce(Al$_{0.95}$Ni$_{0.05}$)$_2$ - the magnetic field values of metamagnetic transition (see text).

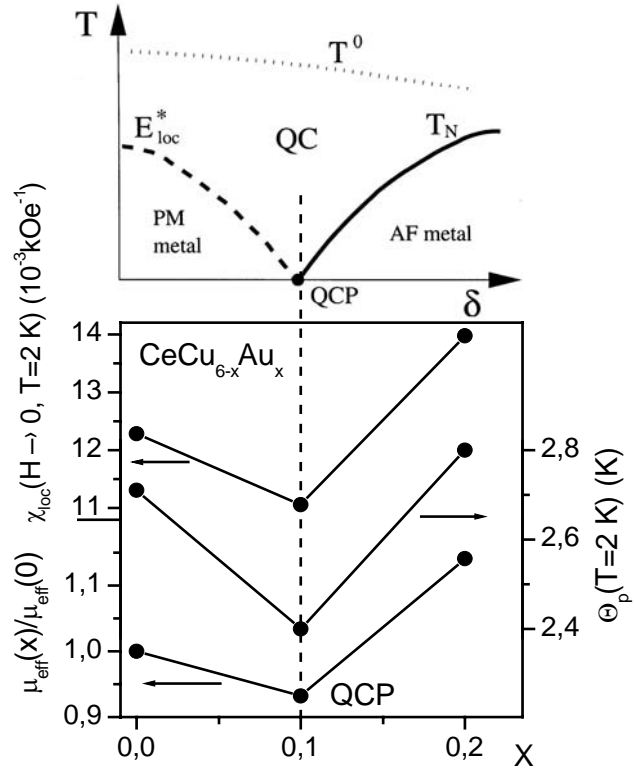

**Fig.9.** Concentration dependencies of local magnetic susceptibility $\chi_{loc}(H\to 0, T=2\ K)=f(x)$ and parameters $\Theta_p(T=2\ K)(x)$ and $\mu_{eff}(x)/\mu(0)$ (see text) in $CeCu_{6-x}Au_x$. A phase diagram, illustrating the origin of quantum critical behavior (QC) at transition from paramagnetic (PM) to antiferromagnetic (AF) metallic phase, is shown in the upper part of the figure. QCP – quantum critical point.

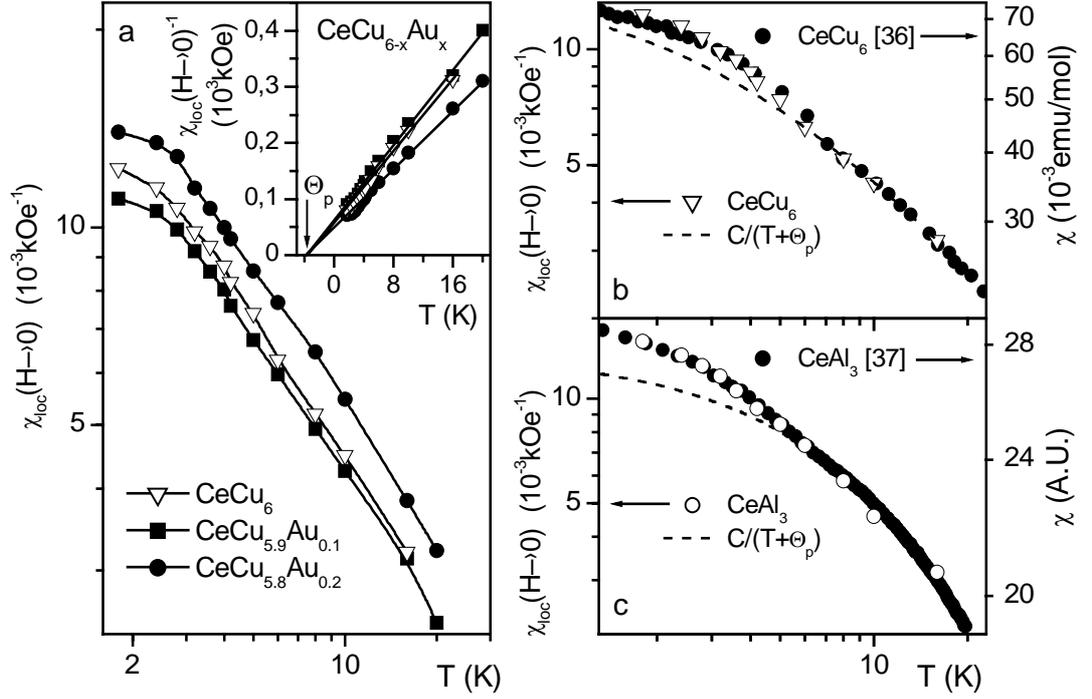

**Fig.10.** Temperature dependencies of local magnetic susceptibility $\chi_{loc}(T,H\to 0)$ (see text) for $CeCu_{6-x}Au_x$ (x=0, 0.1, 0.2) ($\chi_{loc}^{-1}(T,H\to 0)$ for $CeCu_{6-x}Au_x$ dependencies in Curie-Weiss coordinates are shown in the inset (keys preserved)) (**a**). Data of $\chi_{loc}(T,H\to 0)$ for $CeCu_6$ (**b**) and $CeAl_3$ (**c**) is given in comparison with the results of direct magnetic susceptibility measurements [36, 37].

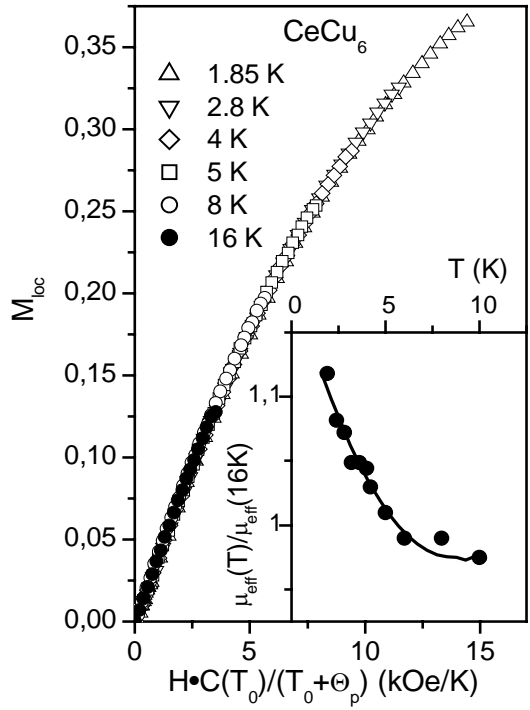

**Fig.11.** $M_{loc}$ dependencies of CeCu$_6$ as a function of parameter $H \cdot C(T_0)/(T_0+\Theta_p)$ for temperatures 1.85 K $\leq T \leq$ 16 K. Inset shows temperature dependence of parameter $\mu_{eff}(T)/\mu(16\ K)$ (see text).

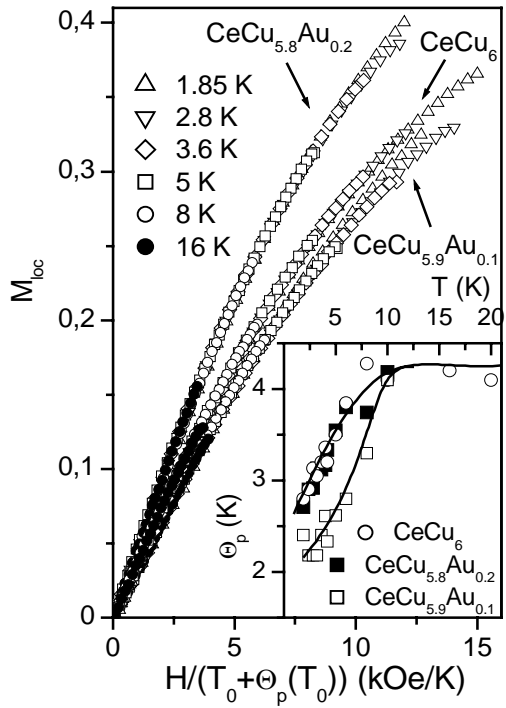

**Fig.12.** $M_{loc}$ dependencies of $CeCu_{6-x}Au_x$ (x=0, 0.1, 0.2) as a function of parameter $H/(T_0+\Theta_p(T_0))$ for temperatures $1.85\ K \leq T \leq 16\ K$. Inset shows temperature dependency of parameter $\Theta_p(T)$ (see text).

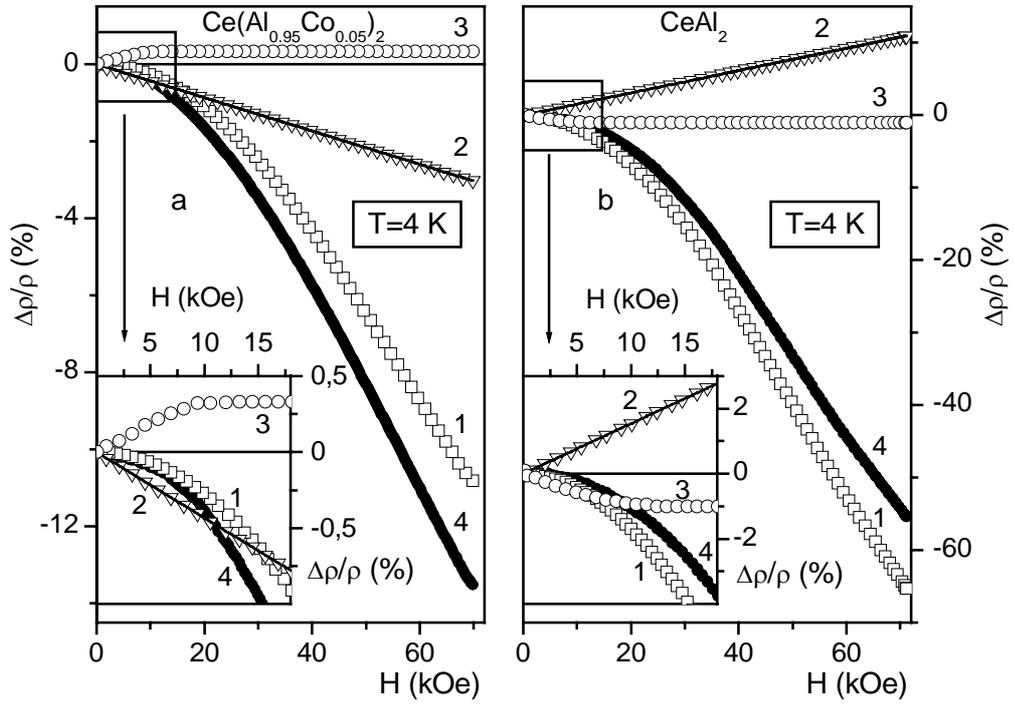

**Fig.13.** Separation of contributions to magnetoresistance $\Delta\rho/\rho$ Ce(Al$_{0.95}$Co$_{0.05}$)$_2$ (**a**) and CeAl$_2$ (**b**) (see text): 1 – "Brillouin-type" quadratic term $\Delta\rho/\rho \sim H^2$, 2 – linear $\Delta\rho/\rho \approx AH$, 3 – low-field magnetic contribution $\Delta\rho/\rho|_{mag}$, 4 – experimental data (see text).

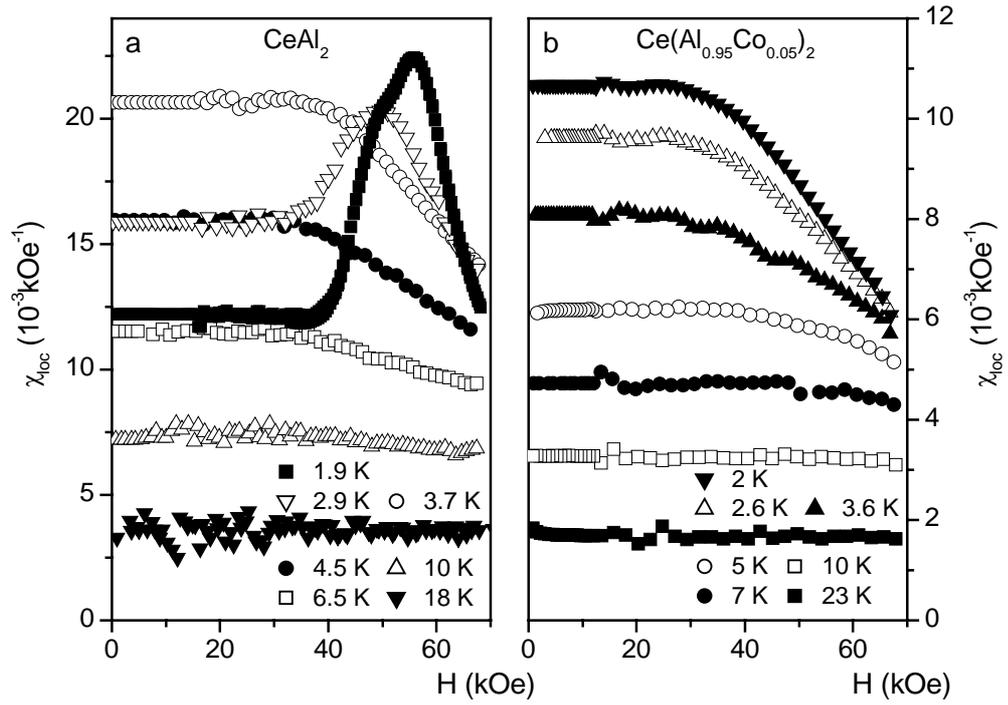

**Fig.14.** Field dependencies of local magnetic susceptibility $\chi_{loc}(H,T_0)$ CeAl$_2$ (**a**) and Ce(Al$_{0.95}$Co$_{0.05}$)$_2$ (**b**) for temperatures 1.9 K $\leq T \leq$ 23 K.

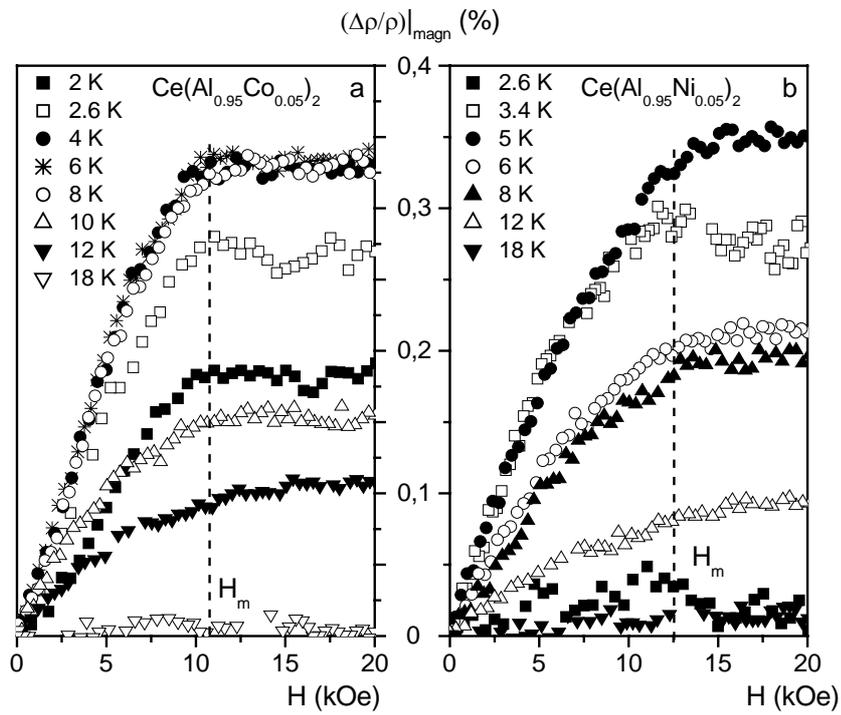

**Fig.15.** Field dependencies of low-field magnetic contribution $\Delta\rho/\rho|_{mag}=\beta\, m_{loc}^2$ to magnetoresistance of Ce(Al$_{0.95}$Co$_{0.05}$)$_2$ (**a**) and Ce(Al$_{0.95}$Ni$_{0.05}$)$_2$ (**b**) for temperaturs $2\,K \leq T \leq 18\,K$. Values $H_m \approx 11$ kOe for Ce(Al$_{0.95}$Co$_{0.05}$)$_2$ (**a**) and $H_m \approx 12.5$ kOe for Ce(Al$_{0.95}$Ni$_{0.05}$)$_2$ (**b**) correspond to metamagnetic phase transition field.

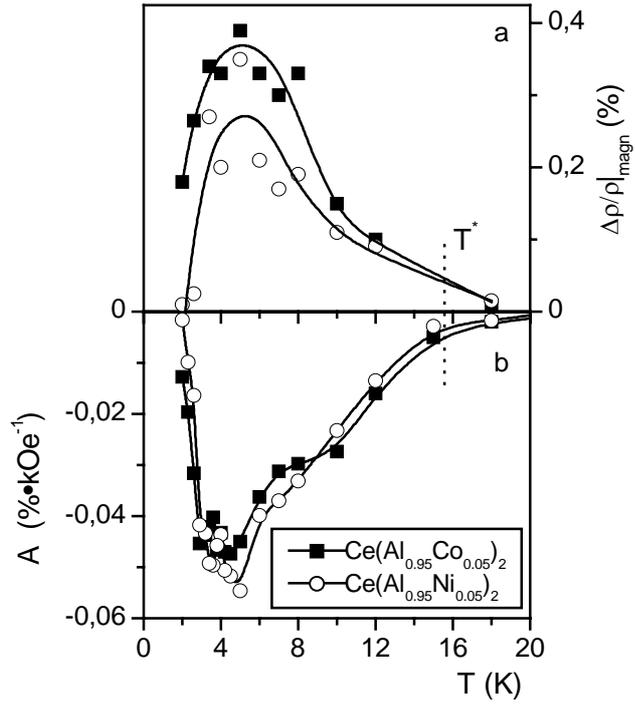

**Fig.16.** Temperature dependencies of the low-field magnetic contribution $\Delta\rho/\rho|_{mag}=\beta\, m_{loc}^2$ at $H>12.5$ kOe (value of saturation) (**a**) and of the parameter $A=2\beta\, \chi_{loc}\, m_{loc}$, corresponding to the linear contribution $\Delta\rho/\rho\approx AH$ (**b**) (see text) to magnetoresistance $\Delta\rho/\rho$ of $Ce(Al_{0.95}M_{0.05})_2$ (M=Co, Ni).

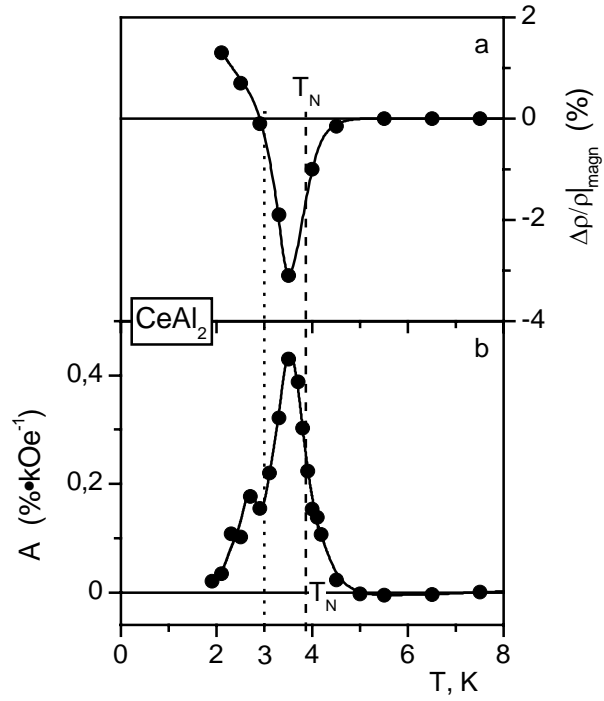

**Fig.17.** Temperature dependencies of the low-field magnetic contribution $\Delta\rho/\rho|_{mag}=\beta\, m_{loc}^2$ at $H>12$ kOe (value of saturation) (**a**) and of the parameter $A=2\beta\, \chi_{loc}\, m_{loc}$, corresponding to the linear contribution $\Delta\rho/\rho \approx AH$ (**b**) to magnetoresistance $\Delta\rho/\rho$ of CeAl$_2$.

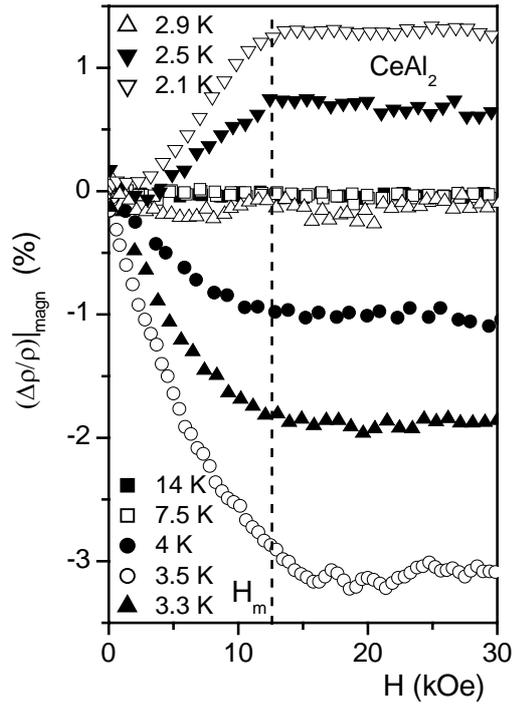

**Fig.18.** Field dependencies of low-field magnetic contribution $\Delta\rho/\rho|_{mag}$ to magnetoresistance of CeAl$_2$ for temperaturs 2.1 K $\leq T \leq$ 14 K. $H_m \approx$ 12 kOe – magnetic field value of metamagnetic transition.

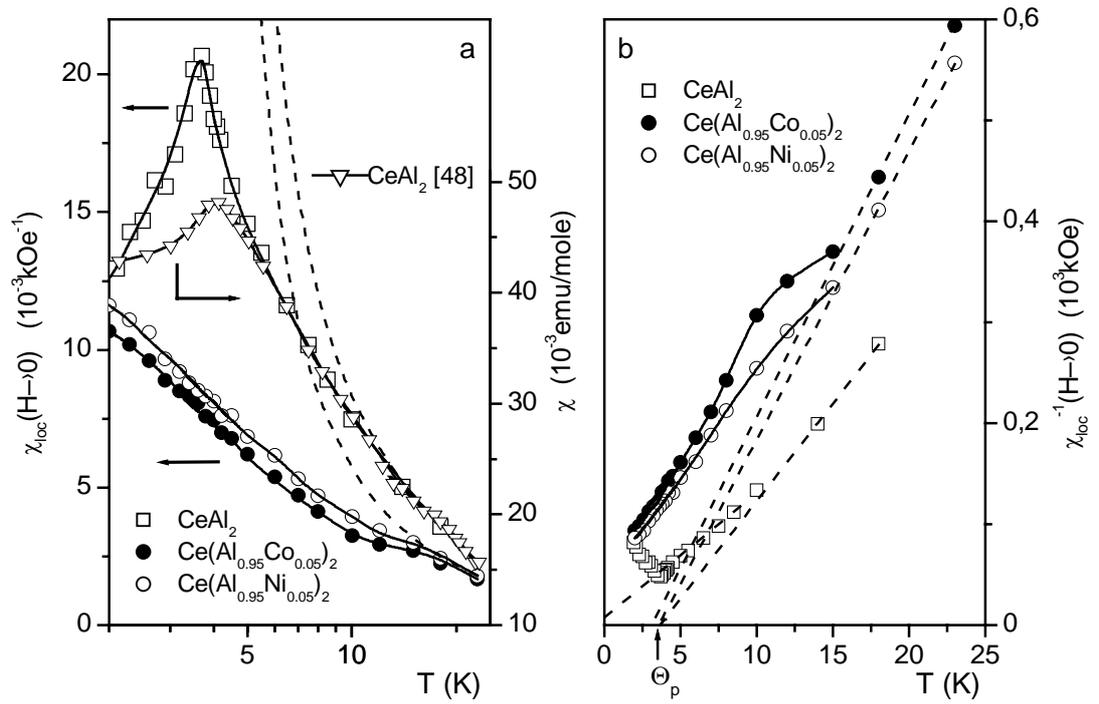

**Fig.19.** Temperature dependencies of local magnetic susceptibility $\chi_{loc}(T,H\to 0)$ (left axis) in comparison to bulk magnetic susceptibility $\chi(T)$ [48] (right axis) for CeAl$_2$ (**a**) and reverse local magnetic susceptibility $\chi_{loc}^{-1}(T,H\to 0)$ of Ce(Al$_{1-x}$M$_x$)$_2$ (x=0, 0.05; M=Co, Ni) (**b**). Dotted line represents Curie-Weiss dependencies $\chi_{loc}^{-1}\sim(T+\Theta_p)$.

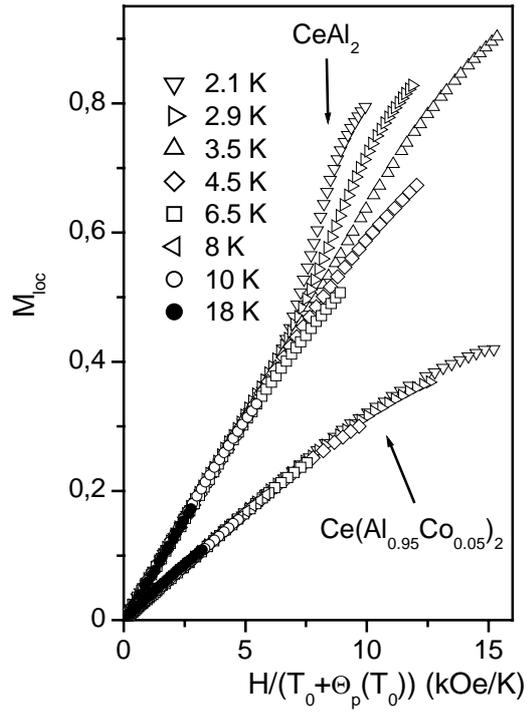

**Fig.20.** $M_{loc}$ dependencies of CeAl$_2$ and Ce(Al$_{0.95}$Co$_{0.05}$)$_2$ as a function of parameter $H/(T_0+\Theta_p(T_0))$ for temperatures 2.1 K $\leq T \leq$ 16 K (see text).

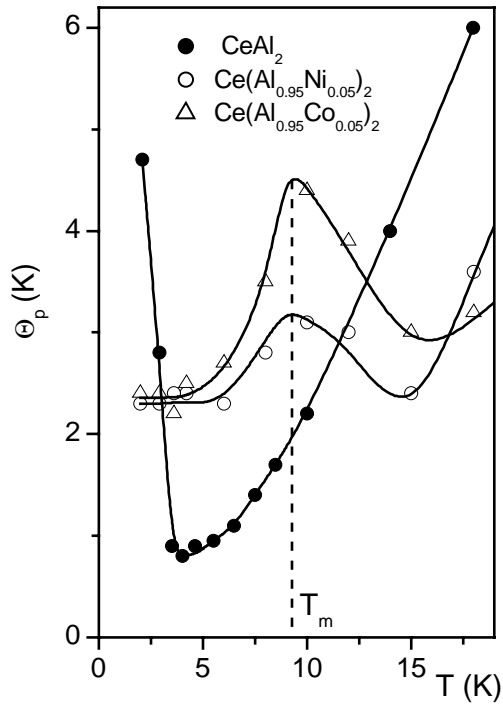

**Fig.21.** Temperature dependencies of $\Theta_p(T)$ parameter for Ce(Al$_{1-x}$M$_x$)$_2$ (x=0, 0.05; M=Co, Ni) (see text).